\documentclass[11pt,reprint,english,aps,prc,showpacs,letter,longbibliography]{revtex4-1}

\usepackage[     bookmarks,
                 bookmarksopen = true,
                 bookmarksnumbered = true,
                 linktocpage,
                 colorlinks = true,
                 linkcolor = blue,
                 urlcolor  = blue,
                 citecolor = blue,
                 anchorcolor = green,
                 hyperindex = true,
                 hyperfigures]
        {hyperref}

\usepackage{graphicx}
\usepackage{dcolumn}
\usepackage{bm}
\usepackage{float}
\usepackage{amsmath}
\usepackage{soul}

\usepackage{color}
\usepackage{graphicx}
\definecolor{dgreen}{cmyk}{1.,0.,1.,0.2}        
\definecolor{orange}{cmyk}{0.,0.353,1.,0.}    

\newcommand\sect[1]{\section{#1}}

\begin{document}

\title{Effects of Initial Nucleon-Nucleon Correlations on  Light Nuclei Production in Au+Au Collisions at $\sqrt{s_\mathrm{NN}} = 3\ $ GeV}%

\author{Qian-Ru Lin$^{1}$}
\author{Yu-Jing Huang$^{1}$}
\author{Long-Gang Pang$^{1}$}
\email[]{lgpang@ccnu.edu.cn}
\author{Xiaofeng Luo$^{1}$}
\email[]{xfluo@ccnu.edu.cn}
\author{Xin-Nian Wang$^{1}$}
\email[]{xnwang@ccnu.edu.cn}

\address{$^1$ Key Laboratory of Quark and Lepton Physics (MOE) and Institute of Particle Physics, Central China Normal University, Wuhan 430079, China}

\date{\today}%

\begin{abstract} 
Light nuclei production in heavy-ion collisions serves as a sensitive probe of the QCD phase structure. In coalescence models, triton ($N_t$) and deuteron ($N_d$) yields depend on the spatial separation of nucleon pairs ($\Delta r$) in Wigner functions, yet the impact of initial two-nucleon correlations $\rho(\Delta r)$ remains underexplored. We develop a method to sample nucleons in $^{197}$Au nuclei that simultaneously satisfies both the single-particle distribution $f(r)$ and the two-nucleon correlation $\rho(\Delta r)$. Using these nuclei, we simulate Au+Au collisions at $\sqrt{s_\mathrm{NN}}=3$ GeV via the SMASH transport model (mean-field mode) to calculate proton, deuteron, and triton yields. Simulations reveal a 36\% enhancement in mid-rapidity deuteron yields across all centrality ranges and a 33\% rise in mid-rapidity triton production for 0-10\% central collisions. Calculated transverse momentum of light nuclei aligns with STAR data. We further analyze impacts of baryon conservation, spectator exclusion, and centrality determination via charged multiplicity. Notably, observed discrepancies in the double yield ratio suggest unaccounted physical mechanisms, such as critical fluctuations or inaccuracies in coalescence parameters or light nuclei cross-sections. This underscores the critical role of initial nucleon-nucleon correlations, linking microscopic nuclear structure to intermediate-energy collision dynamics.  
\end{abstract}

\maketitle{}

\sect{Introduction}

The exploration of nuclear matter under extreme conditions is a cornerstone of heavy-ion collision research, with the investigation of phase transitions and the properties of neutron stars at the forefront \cite{Elfner:2022iae,Huth:2021bsp,Cunqueiro:2021wls,Busza:2018rrf,Baym:2017whm,Lattimer:2015nhk}. This paper specifically aims to elucidate the role of nucleon-nucleon correlation in the production of light nuclei, a key to understanding the critical end point (CEP) of the QCD phase diagram \cite{STAR:2021mii,STAR:2022fan}.

Recent advancements have been marked by the publication of data from the RHIC-STAR Beam Energy Scan experiment for Au+Au collisions at $\sqrt{s_{NN}}$ = 3 GeV. The observables measured, including high order cumulants and collective flow, are pivotal for probing the CEP \cite{STAR:2020hya,STAR:2021iop,STAR:2021fge,Chen:2024aom,Luo:2022mtp}. Theoretical predictions suggest a pronounced local baryon density fluctuation near the CEP regime, which is where light nuclei, with their sensitivity to neutron relative density fluctuations, provide a unique probe \cite{Ye:2020lrc,Bzdak:2019pkr, Luo:2020pef, Oliinychenko:2020ply}.

The yield ratio of light nuclei, a significant observable in this context, has been the subject of intense scrutiny \cite{Oliinychenko:2020ply,Sun:2020uoj,Zhao:2021dka}. The non-monotonic behavior observed in heavy-ion collisions at RHIC BES energies offers compelling evidence for the existence of a CEP \cite{Sun:2020pjz,STAR:2022hbp,Vovchenko:2022xil}. To build a foundation for forthcoming experiments, theoretical physicists are increasingly employing transport models, such as BUU and QMD, to simulate heavy-ion collisions and scrutinize the yield ratio of light nuclei \cite{TMEP:2022xjg,Bass:1998ca,Bertsch:1984gb,SMASH:2016zqf}.

Despite the prevalence of these models, few account for the impact of short-range correlations (SRC) in the initial state \cite{JeffersonLabHallA:2007lly}. Given the small binding energy of light nuclei, their production is thought to predominantly occur during the kinetic freeze-out stage of heavy-ion collisions \cite{Kachelriess:2023jis,Sun:2018mqq}. The calculation of light nuclei production can be approached in various ways, including statistical hadronization and coalescence models, which are sensitive to the relative distance between nucleon pairs \cite{Xu:2018jff,Cai:2019jtk,Kozhevnikova:2022wms}. Sampling nucleons from the many-body wave function of Au is an exceedingly challenging task. In previous studies, simulations of heavy-ion collisions have relied on sampling nucleons based on the single-nucleon distribution. However, a minimal yet essential correction involves accounting for the correlation between two nucleons. This correlation, often referred to as the two-nucleon relative distance distribution, plays a significant role in modifying the uniformity of nucleon distribution, initial-state fluctuations, and correlations. 

In this article, we aim to explore the impact of the two-nucleon relative distance distribution on the production of light nuclei. This study introduces a novel Monte Carlo sampling technique that respects both the single-nucleon distribution, following a Wood-Saxon distribution function, and the two-nucleon distribution function $\rho(\Delta r)$ derived from ab-initio calculations \cite{Alvioli:2005cz,Alvioli:2009ab}. We explore the potential influence of the two-nucleon distribution within the nucleus on the yields and differential momentum space distributions of protons, deuterons and tritons.

The structure of this paper is as follows: Section \ref{sec:method} outlines the initial nucleon sampling method, the SMASH transport model, and the coalescence method. Section \ref{sec:results} presents the results of the light-nuclei yield ratio and differential yields, comparing calculations with and without nucleon-nucleon correlation to data from the STAR experiment at RHIC. The paper concludes with a summary of findings and their implications for future research.

\sect{Method}
\label{sec:method}

We employ SMASH to simulate Au+Au collisions at $\sqrt{s_{NN}}=3$ GeV, to investigate the impact of nucleon-nucleon correlation and the effect of baryon conservation on various observables related to light nuclei production. 
SMASH(Simulating Many Accelerated Strongly-interacting Hadrons) is a transport model that simulates the free streaming, scattering, resonance decay, and formation of relativistic hadrons from heavy-ion collisions\cite{Mohs:2020awg,Bashkanov:2015xsa,SMASH:2016zqf,ParticleDataGroup:2016lqr,Steinberg:2018jvv}. SMASH solves the relativistic Boltzmann equation using the Monte Carlo method,
\begin{equation}
p^{\mu}\partial_{\mu}f_{i}(x,p)+m_{i}F^{\alpha}\partial^{p}_{\alpha}f_{i}(x,p)=C^{i}_{coll}
\label{eq:Boltzmann}
\end{equation}
where $f_{i}(x,p)$ is the phase space distribution and $m_i$ is the mass of particle type $i$. $F^{\alpha}$ represents the external force experienced by individual particles, and $C^{i}_{coll}$ is the collision term \cite{Weil:2016zrk}. In practice, a collision is triggered when the transverse distance between two particles is smaller than the maximum interaction radius, calculated from their total cross section as $d_T < \sqrt{\sigma_{tot} / \pi}$\cite{SMASH:2016zqf,osti_4437568,Bizzarri:1973sp}. In the latest versions of SMASH, elastic and inelastic scattering cross sections are considered for light nuclei production. Specifically, the inelastic channel $\pi d \leftrightarrow \pi n p $ has been identified as important for deuteron production \cite{Oliinychenko:2018ugs,Oliinychenko:2018odl,Oliinychenko:2020znl}.

Given that the colliding energy of our simulation is only 3 GeV, which is relatively low, and the mean field plays a significant role in affecting particle motion, we have incorporated the effect of mean field potentials in SMASH. The Hamilton's equations of motion are thus formulated as follows,
\begin{eqnarray}
\frac{d\vec{r}_i}{dt}&=&\frac{\partial{H_i}}{\partial{\vec{p}_i}}=\frac{\vec{p}_i}{\sqrt{\vec{p}_i^2+m_{eff}^2}} \\
\frac{d\vec{p}_i}{dt}&=&-\frac{\partial{H_i}}{\partial{\vec{r}_i}}=-\frac{\partial{U}}{\partial{\vec{r}_i}}
\label{eq:motion_eq1}
\end{eqnarray}
where the relativistic Hamiltonian is given by,
\begin{equation}
H_{i}=\sqrt{\vec{p}_i^2+m_{eff}^2}+U(\vec{r}_i)
\label{eq:Hamiltonian}
\end{equation}
with the mean-field potential $U(\vec{r}_i)$ taking the form of the Skyrme potential,
\begin{equation}
U(\vec{r}_i)=a{\rho(\vec{r}_i) \over \rho_0}+b\left[{\rho(\vec{r}_i)\over \rho_0}\right]^{\tau}\pm2S_{pot}\frac{I_3}{I}\frac{\rho_{I_3}(\vec{r}_i)}{\rho_0}
\label{eq:Hamiltonian}
\end{equation}
where $a$, $b$ and $\tau$ are Skyrme potential parameters set to $a = -209.2$ MeV, $b = 156.4$ MeV, and $\tau = 1.35$ in the default configuration.  The $\rho$ is the net-baryon density and $\rho_{I_3}$ is the density of relative net-baryon-isospin $I_3/I$, in the comoving frame. The $\rho_{0} = 0.168\ fm^{-3}$ is the saturation density in the core of nucleus. The default value for the symmetry potential is $S_{pot} = 18$ MeV. 

To address the significant fluctuations observed in the calculation of the mean field at a given $\vec{r}$, SMASH employs two optional solutions: the test particle method and event ensembles\cite{Tarasovicova:2024isp}. In this study, we employed the event ensemble approach to compute the mean field potential. An event ensemble consists of multiple independent simulation instances initiated with identical initial conditions, each representing a distinct possible evolution trajectory of the system. The mean field potential $U(\vec{r}_i)$ was derived by averaging over the contributions of particles from all instances within the ensemble. This method improves both statistical precision and computational efficiency by enabling parallel execution of multiple simulations. 

Additionally, we incorporated Pauli-blocking into our simulations. Pauli-blocking ensures that the wave functions of fermions obey the Pauli exclusion principle in quantum statistics, thereby preventing two or more fermions from occupying the same quantum state. By integrating Pauli-blocking, we achieved a more accurate representation of particle distribution and behavior, particularly under conditions of high density or low temperature.

To investigate the effect of nucleon-nucleon correlation, we consider two different initial conditions. The first initial condition uses Au nuclei sampled from a single nucleon distribution known as the Wood-Saxon(WS) distribution function. The second initial condition samples nucleons inside Au nuclei whose single nucleon distribution is the same as the first one, but additionally, the nucleon-pair distances follow a two-nucleon distribution function. The WS distribution is given below, 
\begin{equation}
\begin{aligned}
\rho_A(r)=\frac{\rho_0}{1+\exp{(\frac{r-R_A}{d_A})}}
\label{eq:WS}
\end{aligned}
\end{equation}
where $\rho_0=0.17fm^{-3}$ is the nucleon density, $R_A=6.38fm$ and $d_A=0.535fm$ are used in the current calculation. 

The nucleon-nucleon correlation is described by the distribution of the relative distances between each pair of nucleons, denoted as $\rho(\Delta r)$. This distribution is obtained from experimental measurements. 
To incorporate nucleon-nucleon correlation in Au nuclei, we first sample 3D-coordinates of nucleons using the WS distribution, and then accept or reject the newly sampled nucleon according to $\rho(\Delta r')$, where $\Delta r'$ represents the relative distance between this newly sampled nucleon and all nucleons sampled previously. The detailed method for this sampling procedure can be found in \cite{Alvioli:2009ab} and our recent paper to be published \cite{HYJ:2025}.

SMASH provides an option to initiate the simulation of heavy-ion collisions using a customized nucleus from an external file. After the previous NN correlation sampling, we obtain the position coordinate information of 20,000 gold nuclei. The sampling file is then imported into the SMASH executable's Modi as the input for the coordinate information of the target nuclei and the initial state of the incident nuclei. In the simulation, SMASH reads the positional information of one nucleus and then randomly rotates and repositions the next nucleus. This process allows SMASH to simulate a sufficient number of realistic heavy-ion collisions, even if the number of sampled events is limited or if the same coordinate file is used repeatedly.

This paper mainly discusses two forms of light nuclei production,
(i) The first is the doublet resonance state implemented in the SMASH model. The formation of deuteron in the SMASH model primarily occurs through the following four channels:
$p n N \leftrightarrow d N$, $p n \bar{N} \leftrightarrow d \bar{N}$, $p n \pi \leftrightarrow d \pi$, $N N \leftrightarrow d \pi$. However, due to the current limitations in the SMASH model, which does not include multi-particle interactions, this $3\leftrightarrow2$ process is split into $2\leftrightarrow 1$ and $2\leftrightarrow2$ channels, i.e. $p n N \leftrightarrow d' N \leftrightarrow d N$, $p n \bar{N} \leftrightarrow d' \bar{N} \leftrightarrow d \bar{N}$ and $p n \pi \leftrightarrow d' \pi \leftrightarrow d \pi$. Here a fictitious dibaryon resonance $d'$ is used in place of the intermediate state\cite{Oliinychenko:2018odl,Longacre:2013apa}. 
The SMASH model(version smash-SMASH-2.0) we use does not yet consider the production of triton.
It should be mentioned that the latest version of the SMASH model can now produce triton directly, but its running speed is too slow to accumulate enough statistics.
(ii) The second one is the coalescence model where the production probability of a deuteron or triton can be calculated from the Wigner function $f_A$\cite{Zhao:2021dka,Zhao:2020irc,Chen:2003qj,Chen:2003ava,Ko:2012lhi},
\begin{equation}
\begin{aligned}
\frac{\rm{d} N_A}{\rm{d}^3 \textbf{P}_A} &= \frac{g_A}{Z!N!}\int \Pi^Z_{i=1}p^\mu_i\rm{d}^3\sigma_{i\mu}\frac{\rm{d}^3 \textbf{p}_i}{E_i}\textit{f}_{p/\Bar{p}}(\textbf{x}_i,\textbf{p}_i,t_i)\\
&\times \int \Pi^N_{j=1}p^\mu_j\rm{d}^3\sigma_{j\mu}\frac{\rm{d}^3 \textbf{p}_j}{E_j}\textit{f}_{n/\Bar{n}}(\textbf{x}_j,\textbf{p}_j,t_j)\\
&\times\textit{f}_{A}(\bm{\rho},\lambda, \cdots, \textbf{p}_\rho, \textbf{p}_\lambda, \cdots)\\
&\times \delta^{(3)}(\textbf{P}_A - \sum^Z_{i=1}\textbf{p}_i - \sum^N_{j=1}\textbf{p}_j),
\label{eq:WF}
\end{aligned}
\end{equation}
where the $\textit{f}_{p/\Bar{p}}$ and $\textit{f}_{n/\Bar{n}}$ are the phase space distribution of (anti) protons and (anti) neutrons. $g_A$ is the degeneracy factor of the nucleus of A.

\begin{equation}
\begin{aligned}
f_{A=2}(\bm{\rho},\textbf{p}_\rho)=8 \exp[-\frac{\bm{\rho}^2}{\sigma_{\rho}^2}-\textbf{p}_\rho^2 \sigma_{\rho}^2]
\label{eq:WF_d}
\end{aligned}
\end{equation}

\begin{equation}
\begin{aligned}
f_{A=3}(\bm{\rho},\lambda, \textbf{p}_\rho, \textbf{p}_\lambda)=8^2 \exp[-\frac{\bm{\rho}^2}{\sigma_{\rho}^2}-\frac{\bm{\lambda}^2}{\sigma_{\lambda}^2}-\textbf{p}_\rho^2 \sigma_{\rho}^2-\textbf{p}_\lambda^2 \sigma_{\lambda}^2]
\label{eq:WF_t}
\end{aligned}
\end{equation}
where 
\begin{equation}
\begin{aligned}
&\bm{\rho}=\frac{1}{\sqrt{2}} (\textbf{x}_1'-\textbf{x}_2'),    \bm{p}_\rho=\sqrt{2} \frac{m_2 \bm{p}_1'-m_1 \bm{p}_2'}{m_1+m_2},\\
&\lambda=\sqrt{\frac{2}{3}} (\frac{m_1 \bm{x}_1'+m_2 \bm{x}_2'}{m_1+m_2}-\bm{x}_3')\\
&\bm{p}_\lambda=\sqrt{\frac{3}{2}} \frac{m_3 (\bm{p}_1'+\bm{p}_2')-(m_1+m_2) \bm{p}_3'}{m_1+m_2+m_3}
\label{eq:WF'}
\end{aligned}
\end{equation}

The $f_2$ can be used to compute the yield of deuteron (p+n$\rightarrow$d) and two-body triton through (d+n$\rightarrow$t) while the $f_3$ can be used to compute the yield of three-body triton through (p+n+n$\rightarrow$t). The $\sigma_{\rho}$ in coalescence of two-body triton is set to 2.155 fm\cite{Zhao:2020irc,Scheibl:1998tk}. 

In the previous coalescence method, the influence of relative momentum on the relative position was not taken into account. The coalescence method calculated the overlap probability at a given time instance, $t$, corresponding to kinetic freezeout. However, for particles with a particular relative momentum, they may converge and become closer at a subsequent time $t'$(where $t' > t$). If the momentum directions of two particles are convergent, a minimum distance $d_\perp$ will exist. At this minimum distance, the probability of being coalescence could exceed the initial probability $p$. In appendix B, we compare the results with and without this correction for coalescence.

Following the experimental approach, we first average $N_t$, $N_p$ and $N_d$ before calculating the ratio,
\begin{equation}
\begin{aligned}
\frac{N_t N_p}{N_d^2}=\frac{\langle N_t\rangle \langle N_p\rangle}{\langle N_d\rangle^2}.
\label{eq:particle ratio}
\end{aligned}
\end{equation}

In the traditional coalescence model, net baryon conservation is not considered, leading to a double counting issue where the number of used protons and the number of used deuterons are not subtracted from their yields. 
Recently, a negative Pearson correlation between the measured $\bar{p}$ and $\bar{d}$ yields in Pb+Pb $\sqrt{s_{NN}}=5020$ GeV collisions is observed and explained by the net baryon conservation using a Canonical Ensemble formulation of the statistical hadronization model \cite{ALICE:2022amd}, indicating the importance of removing the double counting.
In our calculation, to avoid the double counting of protons and neutrons, instead of considering 3-body triton production, we utilize SMASH to provide the number of protons and deuterons. We then incorporate two-body triton production (d+n$\rightarrow$t) in the coalescence model. 
To address the double counting, we correct the the $N_d$ by subtracting the number of used deuterons in the two-body triton production.
The number of protons from SMASH does not require correction since the protons used in deuteron production have already been accounted for.
This correction method provides an effective way to incorporate net baryon conservation in our calculations.

The determination of collision centralities was based on the charged-particle multiplicity within the pseudo-rapidity range $0 < \eta < 2$ (FXTMult), considering only those particles with $p_T > 0.4$\cite{STAR:2023uxk}. The $\eta$ range was included to ensure consistency with the experimental data.
Because there is almost no spectator observed in the experiment, we applied a $p_T$ cut to all events to remove the impact of spectators. Based on this method, we obtained four centrality classes: $0-10\%$, $10-20\%$, $20-40\%$and $40-80\%$. The 
total number of events with nucleon-nucleon correlation is 200,000, with each of the four classes of centrality being 19,468, 19,468, 38,936, and 77,872. Similarly, the total number of events without nucleon-nucleon correlation is also 200,000, with 15,898, 15,898, 31,796, and 63,592 events in each of the four centrality classes, respectively. Additionally, the method of centrality classification significantly impacts our results. For details, refer to the appendix C, which shows the charged-particle multiplicity distribution based on impact parameter.

\begin{figure}[htp]
\centering
\includegraphics[width=0.48\textwidth,trim=40 23 75 65,clip]{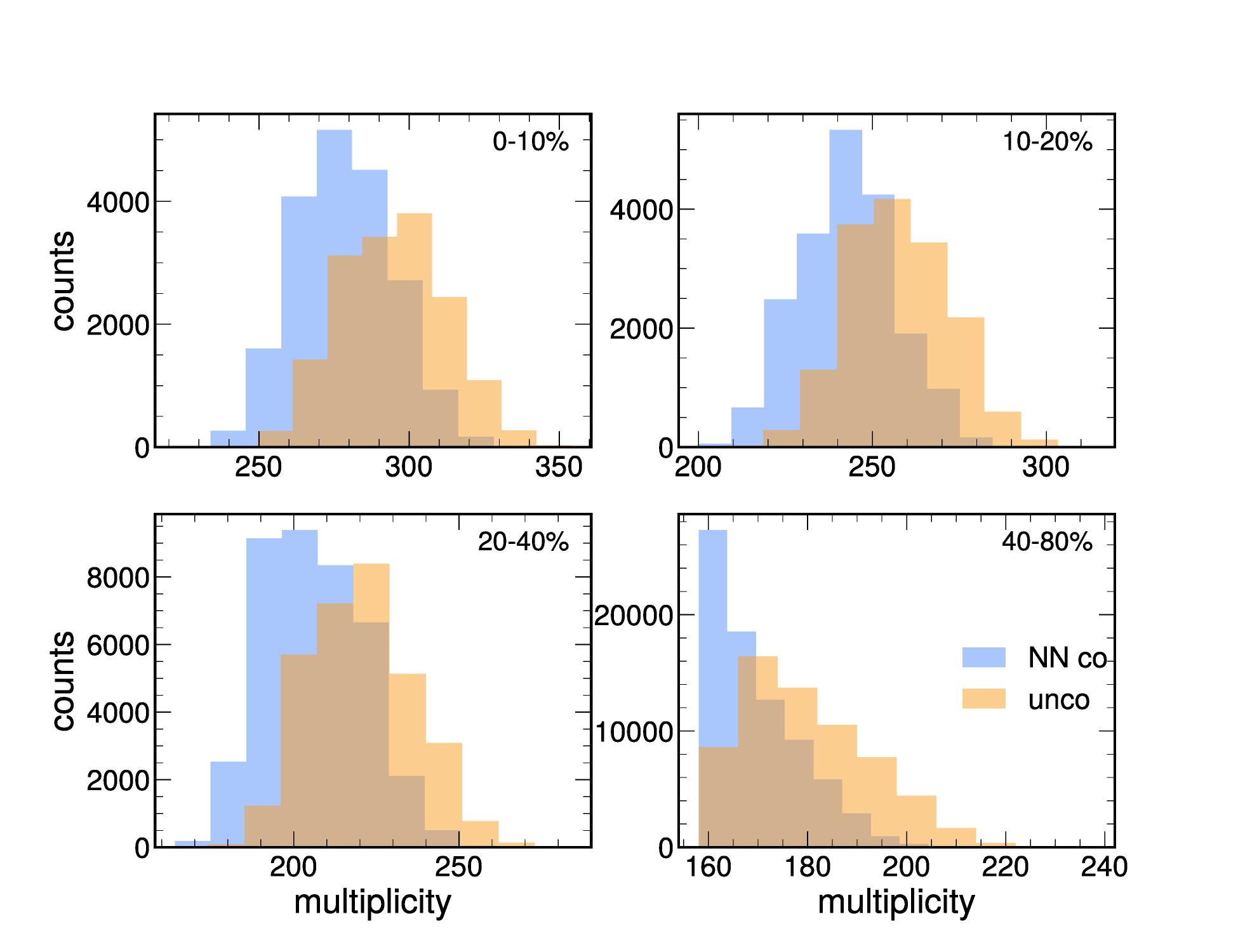}
\caption {(Color online) Distribution of charged-particle multiplicity in four different centrality classes for Au+Au collisions at $\sqrt{s_{NN}}$ = 3 GeV, starting from sampled nucleons in nuclei with (filled blue) and without (filled orange) nucleon-nucleon correlation.}
\label{fig:mutiplicity}
\end{figure}

Figure \ref{fig:mutiplicity} shows the charged-particle multiplicity in four different centrality classes for Au+Au collisions at $\sqrt{s_{NN}}$ = 3 GeV, incorporating the effect of nucleon-nucleon correlation and its absence. The blue area represents the results with correlation included, while the orange area indicates the results without correlation. The distributions of charged-particle multiplicity are observed to be lower with the inclusion of correlation across all four centrality classes, as compared to the distributions without correlation.

\sect{Results}
\label{sec:results}

\begin{figure*}[htp]
\centering
\includegraphics[width=1\textwidth,trim=10 5 8 20,clip]{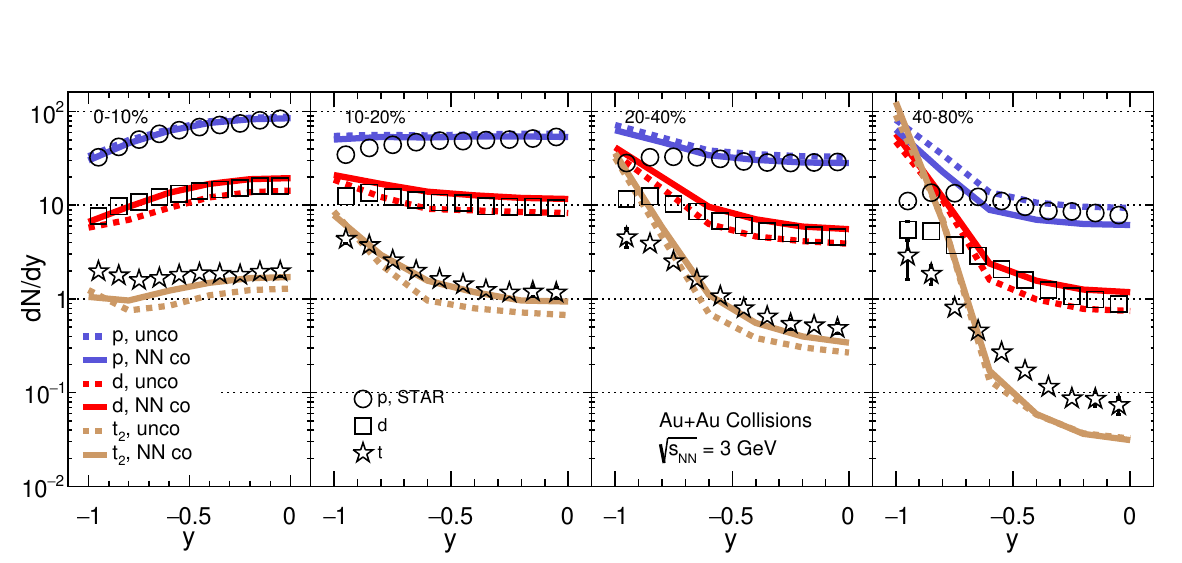}
\caption {(Color online) The yield of light nuclei as a function of rapidity $y$ in 0$\sim$10\%, 10$\sim$20\%, 20$\sim$40\% and 40$\sim$80\% centrality using the SMASH model and coalescence method for Au+Au collisions at $\sqrt{s_{NN}}$ = 3 GeV. The dashed lines are the results for initial nuclear structure without NN correlation in the SMASH simulation, where blue, red and brown one represent proton, deuteron and triton respectively. The solid lines are the results from SMASH with NN correlations in the initial state, where blue one is proton, red one is deuteron and brown one is triton. The dots in each figure are the STAR experimental results, where circle, square and pentagon represent proton, deuteron and triton respectively\cite{10.21468/SciPostPhysProc.10.040,STAR:2023uxk}.}
\label{fig:dndy}
\end{figure*}

Figure~\ref{fig:dndy} illustrates the rapidity distribution of light nuclei yields in comparison with the STAR experiment for Au+Au collisions at $\sqrt{s_{NN}}$ = 3 GeV across four different centrality classes. The results demonstrate that the yields of deuteron and triton are enhances when nucleon-nucleon correlations are incorporated, compared to the scenario without such correlations. In contrast, the yield of proton decreases after the introduction of correlation. Furthermore, as the centrality of collisions increases, the difference in proton yields between the cases with and without correlation also becomes more pronounced. These findings are robust against statistical uncertainties, as discussed in appendix D.

Visually, the triton yields exhibit no discernible difference between scenarios with and without nucleon-nucleon correlations in peripheral collisions (40-80\%), in contrast to central and semi-central collisions. This phenomenon may be attributed to the double coalescence mechanism of tritons, which is highly sensitive to the local densities of three nucleons and neutron-deuteron (n-d) pairs. In peripheral collisions, these densities are significantly reduced, rendering the effects of nucleon-nucleon correlations on triton yields negligible and thus difficult to observe. Conversely, the proton yield appears to undergo minimal changes in central collisions. This is primarily due to the already substantial proton yield in such collisions, where the relative impact of variations in deuteron yield is proportionally minor.
It is noteworthy that in the large rapidity region of peripheral collisions, the model's calculated results exhibit a significant overestimation compared to the experimental values. This discrepancy is attributed to the influence of spectator nucleons. As demonstrated in appendix E, the issue can be effectively resolved by removing a specific number of nucleons from the spectator region.

For deuterons, their formation mechanism primarily involves intermediate resonances. In contrast, tritons are generated through the coalescence model, where the Wigner function is employed to calculate their formation probability. Notably, the production of both deuterons and tritons relies on protons as fundamental constituents. Both the intermediate resonance mechanism and the coalescence model depend critically on the relative distance and relative momentum between nucleon pairs. Consequently, incorporating more realistic correlations in the relative distance between nucleons is expected to significantly influence the yields of final-state particles. It should be emphasized that while the relative momentum between nucleons also plays a crucial role in determining the yields of light nuclei, this factor is not addressed in the current study.

\begin{figure*}[htp]
\centering
\includegraphics[width=1\textwidth,trim=7 3 32 0,clip]{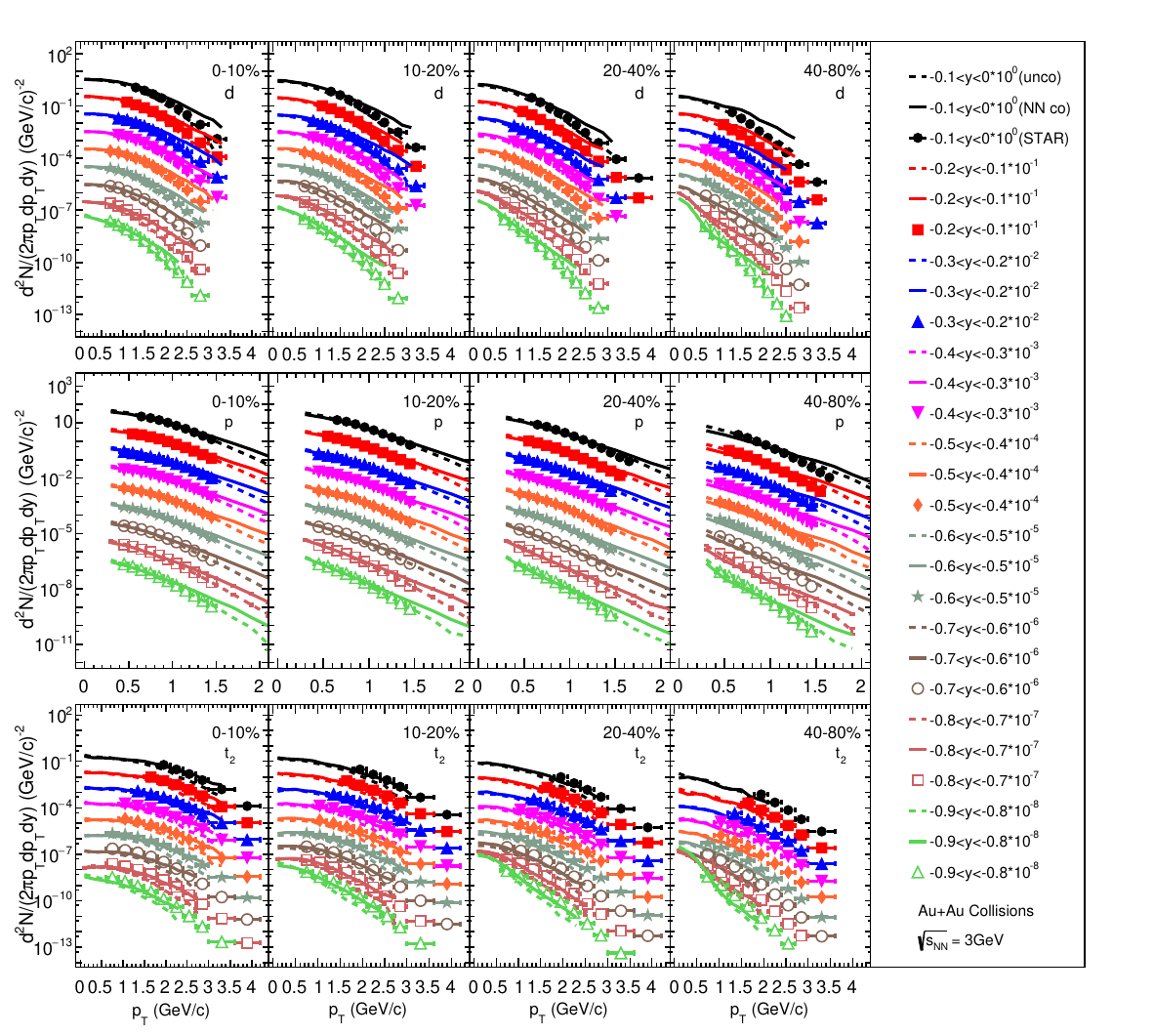}
\caption {(Color online) The transverse momentum($p_T$) spectra of light nuclei from different rapidity windows in 0$\sim$10\%, 10$\sim$20\%, 20$\sim$40\% and 40$\sim$80\% centralities using the SMASH model and coalescence method for Au+Au collisions at $\sqrt{s_{NN}}$ = 3 GeV. The three rows of panels above represent the deuteron, proton, and triton particles in turn. The three panels in each row represent the distribution of the $p_T$ spectra for the four centralities 0-10\%, 10-20\%, 20-40\%, and 40-80\% in turn. In each panel, the dashed line are the results for initial nuclear structure without NN correlation in the SMASH simulation, where the different colours represent different rapidity ranges. The solid lines are the results from SMASH with NN correlations in the initial state, and the dots in each figure are the STAR experimental results \cite{10.21468/SciPostPhysProc.10.040,STAR:2023uxk}.}
\label{fig:pt_spectra}
\end{figure*}

Fig.~\ref{fig:pt_spectra} presents the $p_T$ spectra of light nuclei from different rapidity windows. They are compared with the STAR experimental for Au+Au collisions at $\sqrt{s_{NN}}$ = 3 GeV across four different centralities. The figure is organized into three rows, corresponding to deuteron, proton, and triton particles respectively. Each row contains three panels, which demonstrate the $p_T$ distributions for the centrality ranges 0-10\%, 10-20\%, 20-40\%, and 40-80\%. In each panel, the dashed line represents the results from SMASH simulation without initial nucleon-nucleon (NN) correlation, with different colors indicating various rapidity ranges. with distinct colors denoting different rapidity intervals. The solid lines illustrate the outcomes from SMASH simulations incorporating NN correlations in the initial state, while the data points correspond to the experimental results from the STAR collaboration \cite{10.21468/SciPostPhysProc.10.040}. All curves exhibit a monotonic decrease with increasing transverse momentum. It was observed that the $p_T$ distribution of tritons exhibits an enhancement and aligns more closely with experimental data at low $p_T$, when initial conditions include NN correlations, compared to scenarios without such correlations. Conversely, the production of high $p_T$ protons and deuterons is increased in the presence of initial state NN correlations. These findings suggest that NN correlations significantly influence the momentum distribution of light nuclei.

It should be noted that the experimental $p_T$ spectra have different $p_T$ cutoffs across different rapidity windows due to the limited detector acceptance. Specifically, the minimum detectable $p_T$ for deuteron within the rapidity range $-0.1 < y < 0$ is approximately 1.3 GeV for all centrality classes, and for triton, the minimum detectable $p_T$ is about 1.95 GeV. The $p_T$ acceptance is broader in the forward-rapidity regions than in the middle-rapidity regions. As a result, no experimental data for low $p_T$ regions is available. However, all the figures shown above utilize protons, deuterons and tritons from all $p_T$ regions, both in SMASH simulations and experimental measurement.  In the experimental analysis, the blast wave model is employed to fit the particle yields within the acceptance region, and the model predictions are extended to regions beyond the acceptance. Since the yields of these particles are predominately influenced by the low $p_T$ region, even a small prediction error in this region can lead to a significant discrepancy. 

\begin{figure*}[htp]
\centering
\includegraphics[width=1\textwidth,trim=3 5 13 20,clip]{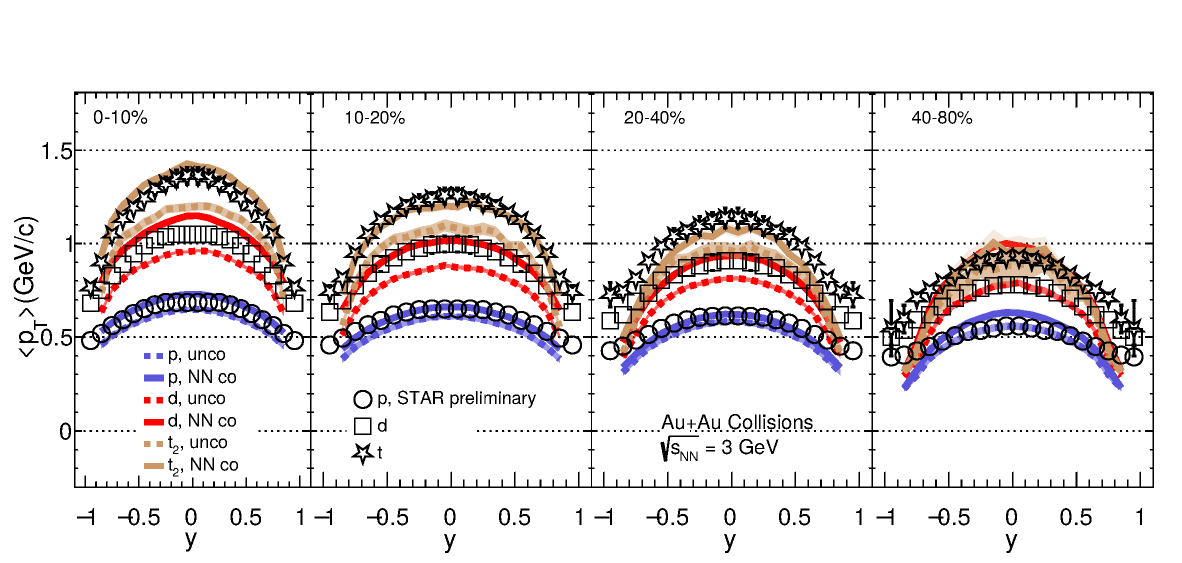}
\caption {(Color online) The mean $p_T$ of light nuclei as a function of rapidity $y$ in 0$\sim$10\%, 10$\sim$20\%, 20$\sim$40\% and 40$\sim$80\% centrality using the SMASH model and coalescence method for Au+Au collisions at $\sqrt{s_{NN}}$ = 3 GeV. The dashed lines are the results for initial nuclear structure without nucleon-nucleon correlation in the SMASH simulation, where green, orange and gray one represent proton, deuteron and triton respectively. The solid lines are the results from SMASH with NN correlations in the initial state, where blue one is proton, red one is deuteron and brown one is triton. The dots in each figure are the STAR experimental results, where circle, square and pentagon represent proton, deuteron and triton respectively\cite{10.21468/SciPostPhysProc.10.040,STAR:2023uxk}.}
\label{fig:mpt}
\end{figure*}

Figure \ref{fig:mpt} illustrates the rapidity distribution of the mean $p_T$ of light nuclei in comparison with the STAR experiment for Au+Au collisions at $\sqrt{s_{NN}}$ = 3 GeV, across four different centrality. 
The results demonstrate that the mean transverse momentum of light nuclei shows better agreement with experimental data when correlation effects are properly accounted for. This observation is consistent with our theoretical predictions, thereby validating the importance of incorporating correlation factors in the analysis framework. 
The remaining discrepancies between SMASH simulations and experimental data for protons, deuterons, and tritons can be attributed to two primary factors.
From an experimental perspective, deviations in yields at low $p_T$ may stem from inconsistencies with the Blast wave model. Our systematic studies reveal that the mean $p_T$ is highly sensitive to the $p_T$ cut in the calculations, where even minor adjustments to the $p_T$ cut can lead to substantial variations in the resulting mean $p_T$. From a theoretical standpoint, the SMASH model may overestimate the production yields of light particles in the large rapidity regions corresponding to spectator. To facilitate more reliable comparisons between theoretical predictions and experimental measurements, we recommend direct comparison of uncorrected experimental data with theoretically calculated results that have undergone appropriate acceptance corrections. This approach would minimize potential biases introduced during the data extrapolation process.

There are also results for transverse momentum from a variety of other models. The results from the UrQMD with thermodynamical approach, as presented in \cite{Kozhevnikova:2024itb,Kozhevnikova:2023mnw}, is in general agreement with the conclusions drawn from our implementation of the SMASH model. In this study, it calculates the yields of light nuclei across various centralities for smooth crossover, first-order phase transition, and hadronization. By comparing these calculated results with experimental data, the study validates that the dynamics of heavy-ion collisions in Au+Au collisions at $\sqrt{s_{NN}}$ = 3GeV are predominantly governed by the hadronic phase. This finding underscores the significance of hadronic interactions in shaping the outcomes of such collisions at this specific energy regime. However, it is important to note that their centrality classification is based on the impact parameters. Additionally, two other studies employing the JAM model combined with the coalescence mechanism also utilize impact parameters for their centrality classification\cite{Xu:2023xul,Liu:2024ilw}.

\begin{figure}[htp]
\centering
\includegraphics[width=0.48\textwidth,trim=6 3 50 45,clip]{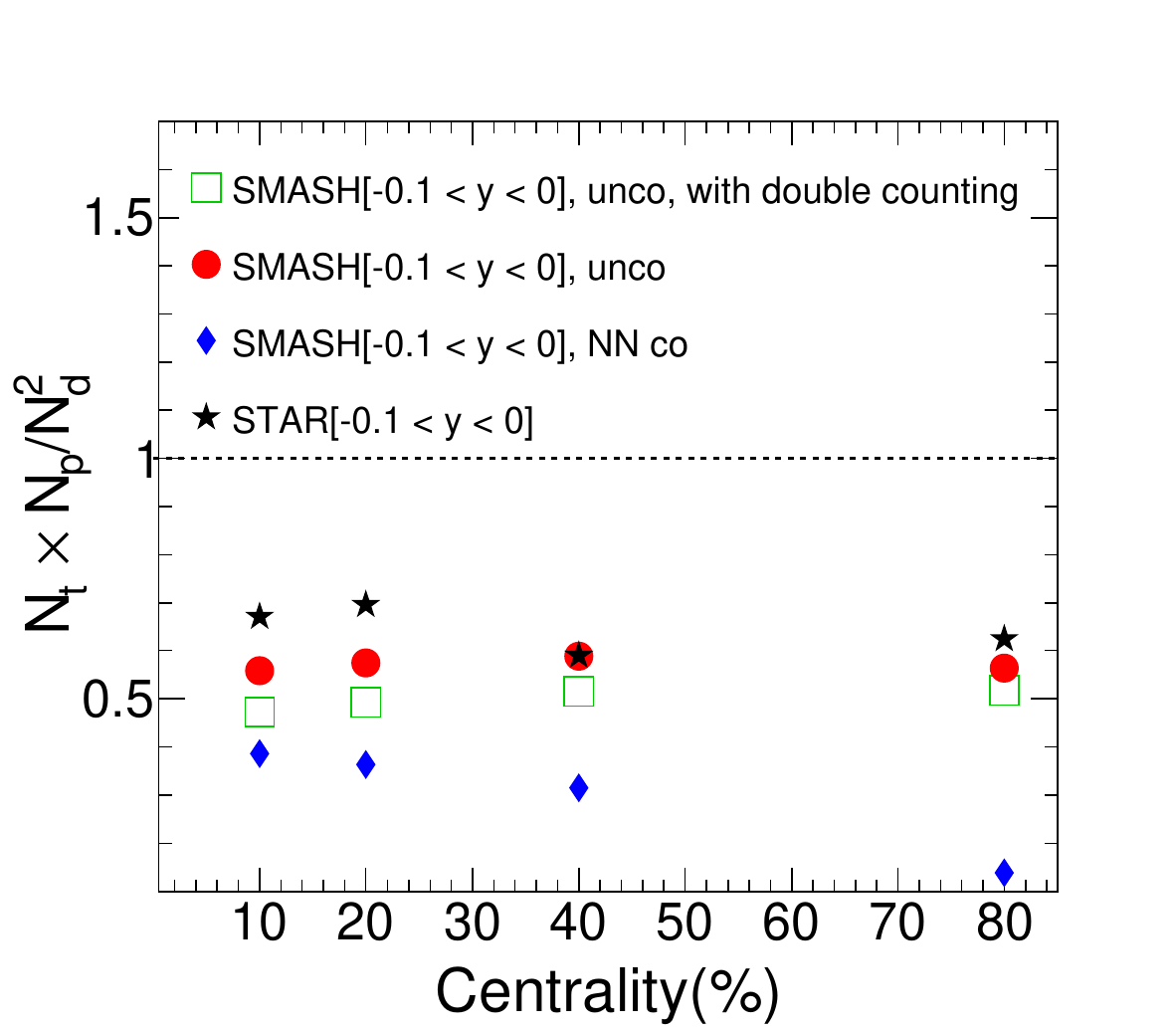}
\caption {(Color online) The yield ratio of $N_t \times N_p/N_d^2$ obtained using the SMASH model for $p$ and $d$  production and $d+n\rightarrow t$ 2-body coalescence method for triton production, starting from sampled nucleons in nuclei with nucleon-nucleon correlation (filled blue-rhombus) and without (filled red-circle), compared with the STAR experiment result, in four different centralities in Au+Au collisions at $\sqrt{s_{NN}}$ = 3GeV \cite{Liu:2022ump,STAR:2023uxk}. As a comparison, the yield ratio with double counting is shown in green square.}
\label{fig:ratio_vs_cent}
\end{figure}

Figure~\ref{fig:ratio_vs_cent} shows our calculated yield ratio of $N_t \times N_p/N_d^2$ using the SMASH model for $p$ and $d$  production and $d+n\rightarrow t$ 2-body coalescence method for triton production, compared with the STAR experimental result in Au+Au collisions at $\sqrt{s_{NN}}$ = 3GeV \cite{Liu:2022ump,10.21468/SciPostPhysProc.10.040}.
Notice that in the present study, we have corrected the deuteron yield by removing the number of deuterons used in triton production. 
Without removing this double counting, the $N_d$ in $N_t \times N_p/N_d^2$ is overestimated, which will lead to a smaller yield ratio than the experimental measurement, as shown by the empty-green squares. 
it is evident that after applying baryon number conservation correction to the deuteron yields, the deuteron yields decrease, thereby increasing the double ratio value and align closer with experimental data.
In contrast, the inclusion of NN correlation significantly reduces the double ratio, which can be attributed to the enhancement of deuteron yield and the reduction of proton yield. The square of the deuteron number exerts a much stronger suppression on this double ratio. However, the incorporation of NN correlation also increases the triton yield and elevates the mean $p_T$ of both deuterons and tritons across various rapidities, resulting in a significantly improved alignment with experimental measurements. Additional physical mechanisms may contribute to the observed discrepancy in the double ratio, such as the influence of initial state momentum and momentum correlations, unaccounted transitions from hadronic to partonic degrees of freedom, the critical fluctuations, discrepancies in the coalescence model parameters or cross-sections related to light nuclei production.

\begin{figure}[htp]
\centering
\includegraphics[width=0.48\textwidth,trim=18 1 55 28,clip]{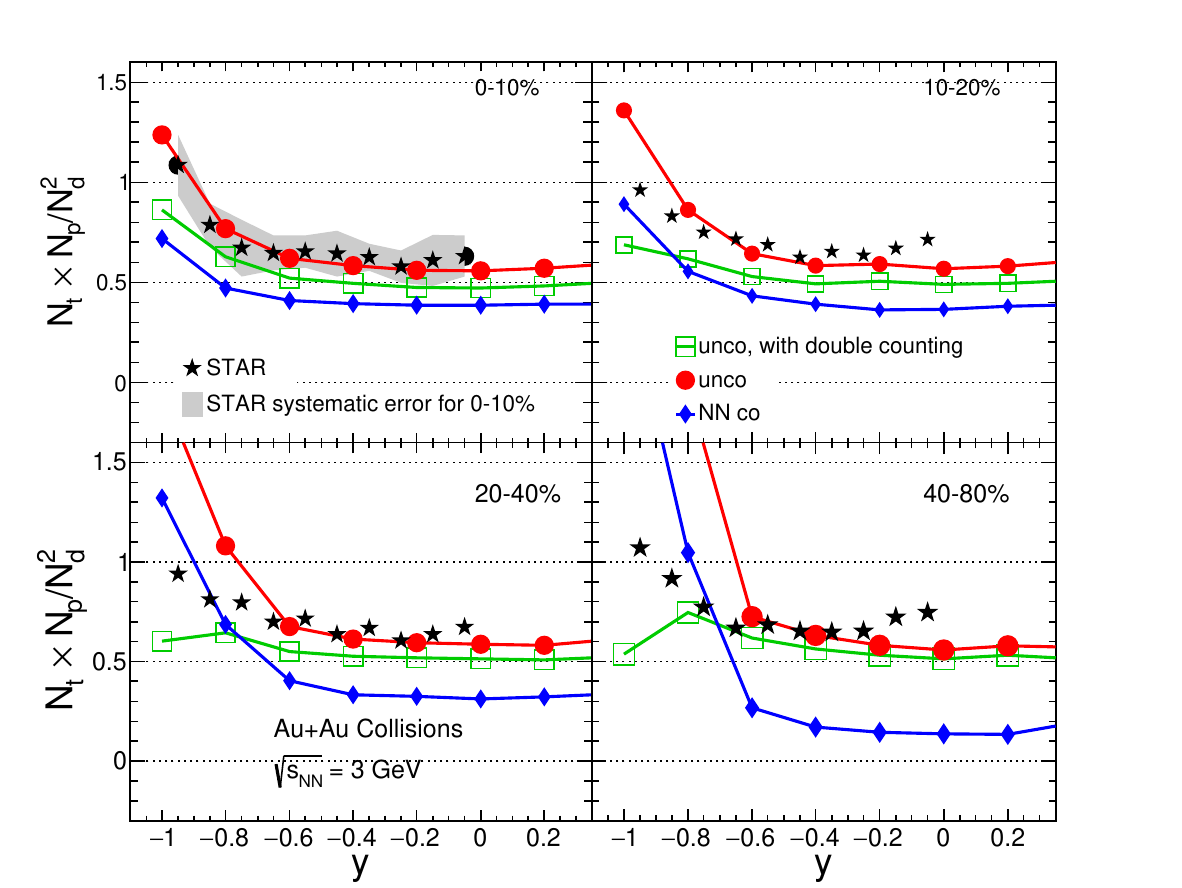}
\caption {(Color online) The yield ratio of $N_t \times N_p/N_d^2$ with respect to rapidity across 0$\sim$10\%, 10$\sim$20\%, 20$\sim$40\% and 40$\sim$80\% centrality, using the SMASH model for $p$ and $d$ production and $d+n\rightarrow t$ 2-body coalescence method for triton production, starting from sampled nucleons in nuclei with nucleon-nucleon correlation (filled blue-rhombus) and without (filled red-circle), compared with the STAR experiment results, in four different centrality classes in Au+Au collisions at $\sqrt{s_{NN}}$ = 3GeV \cite{Liu:2022ump}. For comparison, the yield ratio with double counting is shown in green square.}
\label{fig:ratio_vs_y}
\end{figure}

Figure \ref{fig:ratio_vs_y} illustrates the yield ratio as a function of rapidity for 0$\sim$10\%, 10$\sim$20\%, 20$\sim$40\% and 40$\sim$80\% centrality intervals in Au+Au collisions at $\sqrt{s_{NN}}$ = 3 GeV. The SMASH model was employed for $p$ and $d$  production while the $d+n\rightarrow t$ 2-body coalescence method was used for triton production. Two scenarios, simulations with NN correlation and without, are compared with the STAR experiment result across four different centralities 
The yield ratio obtained from the SMASH model and the 2-body triton coalescence, with double counting corrected and without NN correlation, shows good agreement with the experimental results \cite{Liu:2022ump} for all four centrality intervals. Additionally, the yield ratio calculated with NN correlation is generally lower than that without NN correlation. Qualitatively, the difference is largest in peripheral collisions, which suggests that NN correlations may play a more significant role in these events. 

Both the experimental data and simulations indicate that the yield ratio is higher at backward rapidity $y=-1$ than at middle rapidity $y=0$.
It implies that the yields of light nuclei are affected by the spectator and the target at forward and backward rapidities.
It is important to note that the experimental data have been Lorentz boosted to the center of mass frame. To ensure consistency with the experimental data, the center of mass frame is also used in the simulation of Au+Au collisions in SMASH. In this frame, the rapidity distributions of both experimental data and the simulations are symmetric about $y=0$. This symmetric enhancement further supports the conclusion that the observed yield ratio differences are intrinsic to the collision dynamics rather than artifacts of the fix target experimental setup.


\sect{Summary}
Our study investigated the effect of initial-state nucleon-nucleon (NN) correlations on the production of light nuclei, employing the SMASH model to dynamically generate protons, neutrons, and deuterons, while calculating triton yields using a 2-body coalescence model. The motivation for this work stems from the hypothesis that the Wigner function used in the coalescence model is sensitive to the relative distance between nucleons within the nucleus. Our results demonstrate that incorporating NN correlations in the mean-field mode of SMASH simulations significantly enhances the yields of deuterons and tritons. Additionally, NN correlations increase the mean transverse momentum of these light nuclei, leading to a much better agreement with STAR experimental data.

A notable aspect of our approach is the use of the same method as in experimental studies, where charged multiplicities with a specific momentum cut are used to determine centrality classes. This approach introduces significant corrections. In the deuteron yield calculation, we accounted for baryon conservation by subtracting the deuterons used in the 2-body triton coalescence process. This adjustment reduces deuteron yields and significantly improves the description of the ratio $N_t \times N_p / N_d^2$ when NN correlations are not considered. However, we observed that both our simulations and other transport models fail to describe data at large rapidities. This discrepancy was traced back to the influence of spectator nucleons.

Another technical detail we explored was the production probability of light nuclei, which might be strongest when two nucleons approach each other (but do not collide) with a minimum transverse distance after kinetic freezeout. However, this correction did not yield visible differences in the results.

While NN correlations improve the agreement with experimental data for the yields and mean transverse momentum of deuterons and tritons, they fail to describe the double yield ratio. This is primarily due to the enhanced deuteron yields, as altering the relative distance between nucleons strongly affects light nuclei production. This does not necessarily imply that previous results are superior, as heavy-ion collision simulations are influenced by numerous other initial-state effects and dynamical processes not accounted for in our study, such as the relative momentum between nucleons. Nevertheless, our findings highlight that considering nuclear structures with and without NN correlations in SMASH simulations leads to significant differences in the yield ratios of light nuclei. This confirms that heavy-ion collisions can serve as a valuable tool for studying nuclear structure.

In future work, we plan to explore the effects of the initial momentum distribution and the relative momentum between nucleons on light nuclei production. These investigations could provide further insights into the complex dynamics of heavy-ion collisions and the role of NN correlations in nuclear structure.


\begin{acknowledgments}
We gratefully acknowledge Dr. Hui Liu, Professor Hannah Elfner and Professor Kai-Jia Sun for valuable discussion and suggestions. This work was supported in part by National Natural Science Foundation of China (No.\ 12075098, No.\ 12435009, No.12122505 and No. 12447102, No. 1193507), by the National Key Research and Development Program of
China (No.2022YFA1604900, No.2020YFE0202002), and the Guang-dong MPBAR with No.2020B0301030008.
\end{acknowledgments}

\section{Appendix}

\subsection{Distribution Of Deuteron And Proton Multiplicity From SMASH}

\begin{figure}[htp]
\centering
\includegraphics[width=0.48\textwidth,trim=20 23 75 5,clip]{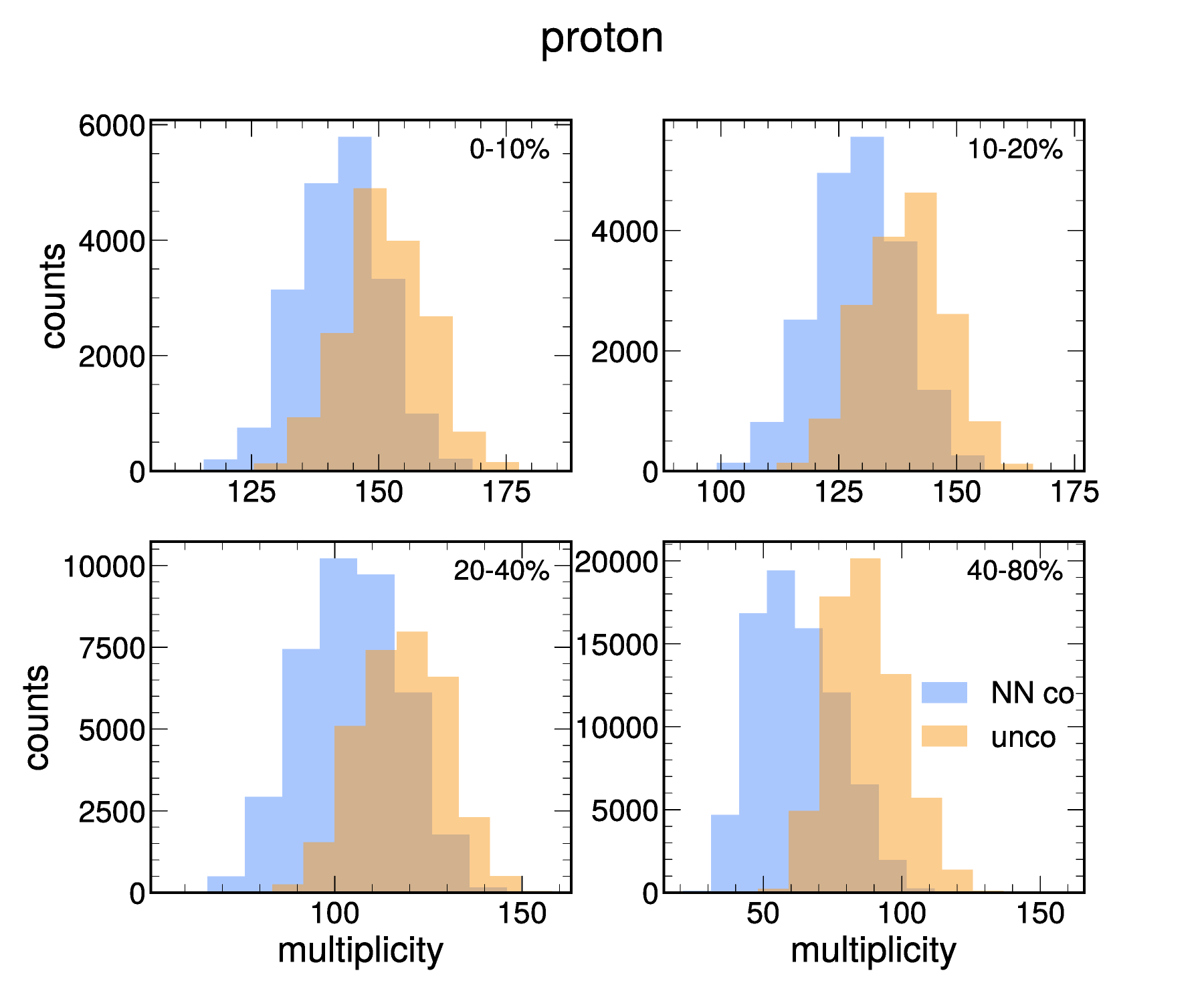}
\caption {(Color online) Distribution of proton multiplicity in four different centrality classes for Au+Au collisions at $\sqrt{s_{NN}}$ = 3 GeV, starting from sampled nucleons in nuclei with (filled blue) and without (filled orange) nucleon-nucleon correlation.}
\label{fig:mutiplicity_proton}
\end{figure}

Figure \ref{fig:mutiplicity_proton} shows the proton multiplicity in four different centrality classes for Au+Au collisions at $\sqrt{s_{NN}}$ = 3 GeV, incorporating the effect of nucleon-nucleon correlation and its absence. The blue area represents the results with correlation included, while the orange area indicates the results without correlation. It can be observed that the proton yield is lower with correlation than without correlation across all four centrality classes.

\begin{figure}[htp]
\centering
\includegraphics[width=0.48\textwidth,trim=20 23 80 5,clip]{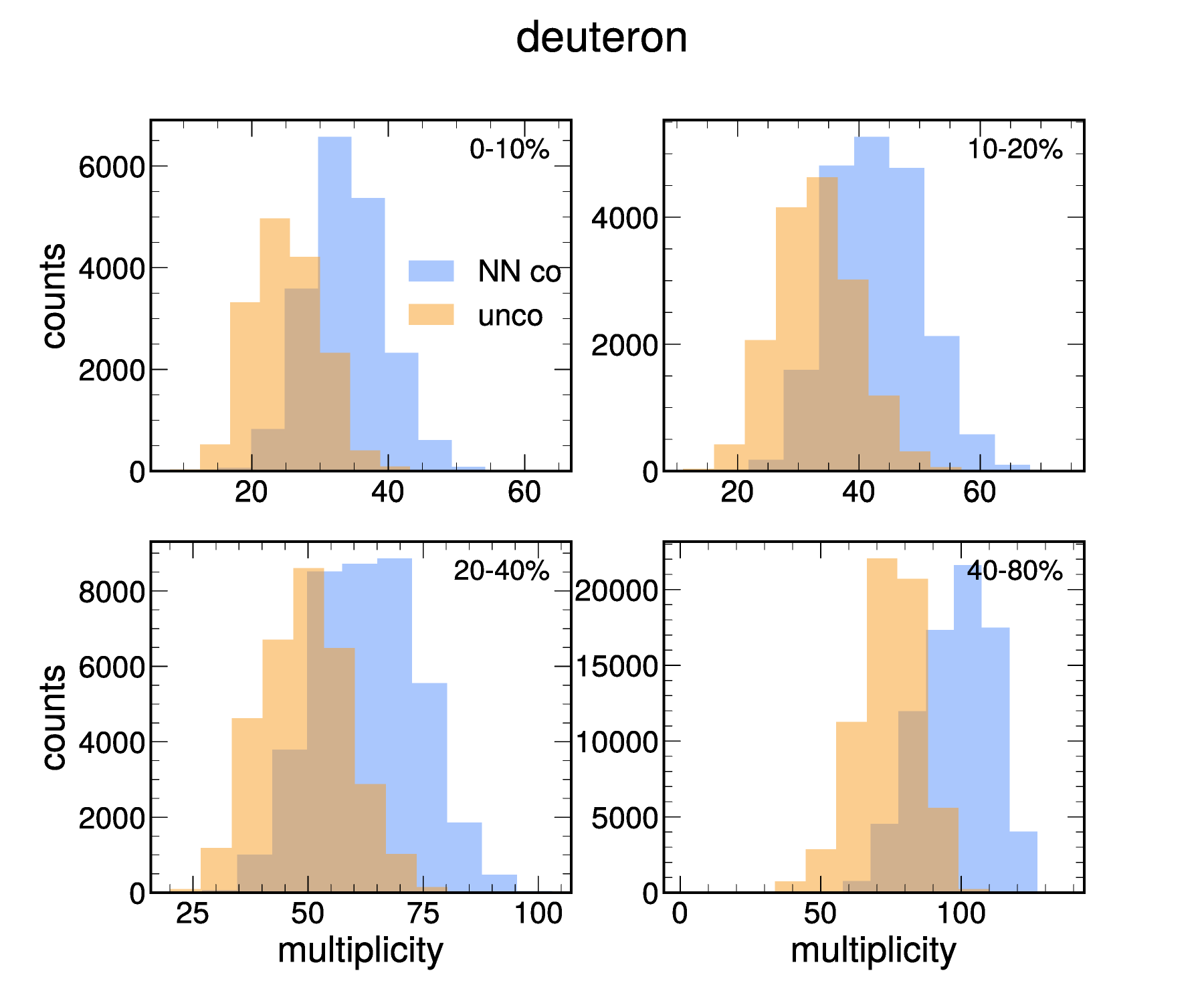}
\caption {(Color online) Distribution of deuteron multiplicity in four different centrality classes for Au+Au collisions at $\sqrt{s_{NN}}$ = 3 GeV, starting from sampled nucleons in nuclei with (filled blue) and without (filled orange) nucleon-nucleon correlation.}
\label{fig:mutiplicity_deuteron}
\end{figure}

Figure \ref{fig:mutiplicity_deuteron} shows the deuteron multiplicity in four different centrality classes for Au+Au collisions at $\sqrt{s_{NN}}$ = 3 GeV, incorporating the effect of nucleon-nucleon correlation and its absence. The blue area represents the results with correlation included, while the orange area indicates the results without correlation. It can be observed that the deuteron yield is lower without correlation than with correlation across all four centrality classes, which is in complete contrast to the results for proton.

\subsection{Coalescence Correction}

\begin{figure}[htp]
\centering
\includegraphics[width=0.5\textwidth,trim=100 50 120 65,clip]{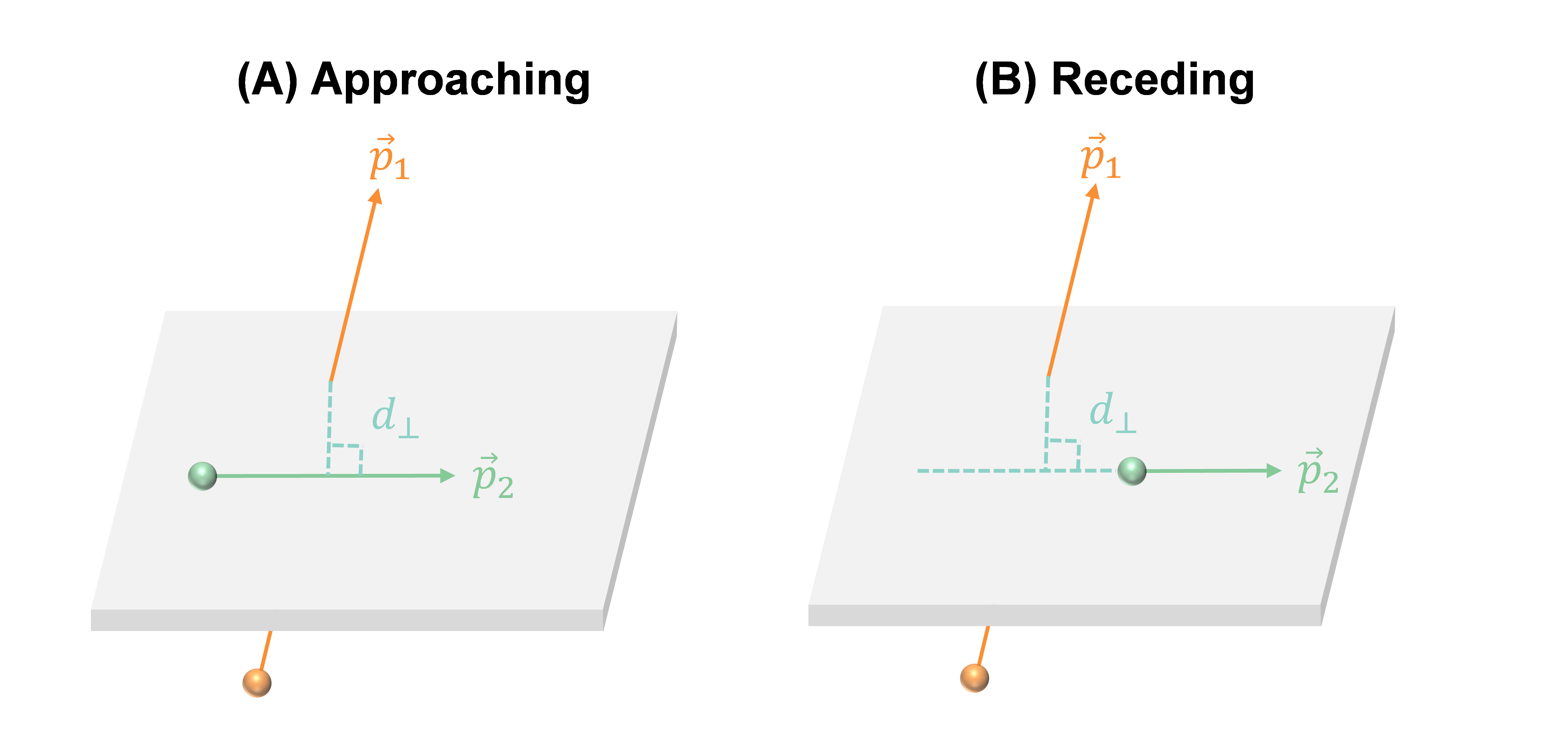}
\caption {(Color online) The schematic diagram of approaching and receding.}
\label{fig:coal}
\end{figure}

Figure \ref{fig:coal} illustrates two scenarios for two-body coalescence after the time of kinetic freeze-out, employing Wigner functions. The first scenario, depicted in Figure \ref{fig:coal} A, represents two nucleons approaching each other, where their initial relative distance is larger than their minimum transverse distance in the near future. In contrast, Fig.~\ref{fig:coal} B presents a receding scenario, where the relative distance might already be the shortest distance in the near future.
The conventional coalescence model typically utilizes the second scenario. However, in scenario A, after the two particles propagate along a straight line, there will inevitably be a minimum distance $d_{\perp}$ and the coalescence probability will be higher than the probability at the initial kinetic freeze-out. The formula for $d_{\perp}$ is as follows: 
\begin{equation}
\begin{aligned}
d_\perp=\sqrt{(\vec{r_1}-\vec{r_2})^2-\frac{[(\vec{r_1}-\vec{r_2})\cdot(\vec{p_1}-\vec{p_2})]^2}{(\vec{p_1}-\vec{p_2})^2}}
\label{eq:dperp}
\end{aligned}
\end{equation}
where $\vec{r}$ and $\vec{p}$ are the coordinates and momenta of the two particles 1 and 2 in the center of mass frame of the binary collision\cite{SMASH:2016zqf}.

\begin{figure*}[htp]
\centering
\includegraphics[width=1\textwidth,trim=10 5 8 20,clip]{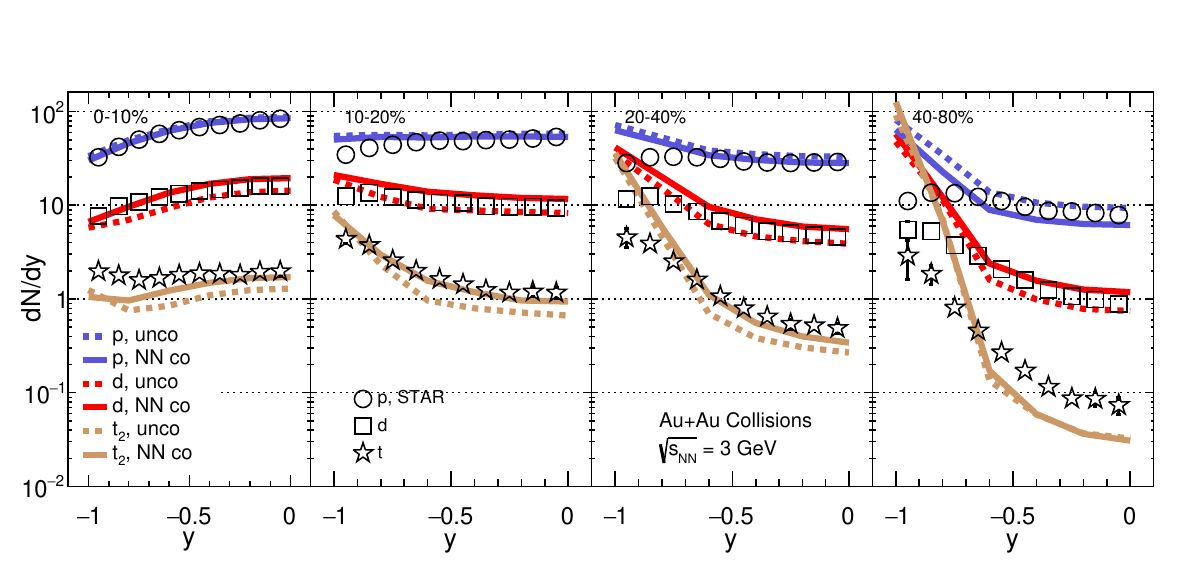}

\caption {Same as Fig.\ref{fig:dndy}, but without coalescence corrections.}
\label{fig:dndy_dignose}
\end{figure*}

Figure \ref{fig:dndy_dignose} presents the results of production of light nuclei obtained from the coalescence method, which only considers the positional convergence probability of two particles and does not take into account the influence of their momentum directions. It can be observed that there is almost no difference when compared to Fig.~\ref{fig:dndy}, which includes the consideration of momentum directions.

\subsection{The Impact of Centrality Classification Methods}

\begin{figure}[htp]
\centering
\includegraphics[width=0.48\textwidth,trim=30 23 80 65,clip]{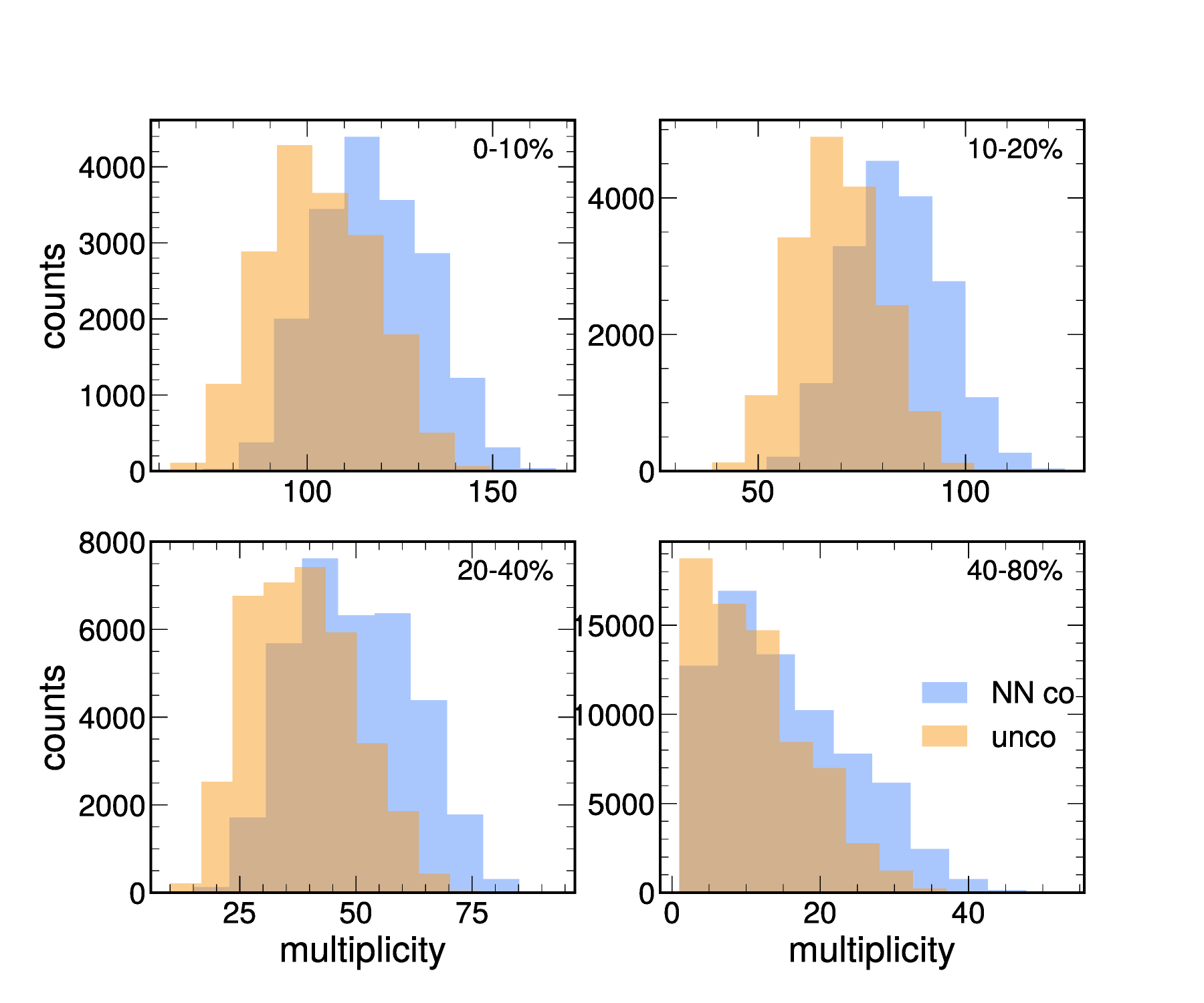}
\caption {(Color online) Distribution of charged-particle multiplicity in four different centrality classes for Au+Au collisions at $\sqrt{s_{NN}}$ = 3 GeV, starting from sampled nucleons in nuclei with (filled blue) and without (filled orange) nucleon-nucleon correlation. Centrality is determined by impact parameter.}
\label{fig:cen_by_b}
\end{figure}

Figure \ref{fig:cen_by_b} shows the distribution of charged-particle multiplicity in four different centrality classes determined by impact parameters,  for Au+Au collisions at $\sqrt{s_{NN}}$ = 3 GeV, incorporating the effect of nucleon-nucleon correlation and its absence. The blue area represents the results with correlation included, while the orange area indicates the results without correlation. This figure differs from the charged-particle multiplicity distribution in the main text, where centrality is determined by the number of final-state charged particles. Here, centrality is determined by the impact parameter. It is evident that these two methods of centrality classification result in noticeable differences in the outcomes. In the centrality results classified by impact parameter, a Gaussian-like multiplicity distribution is obtained, which is evidently unreasonable.

\begin{figure*}[htp]
\centering
\includegraphics[width=1\textwidth,trim=10 5 8 20,clip]{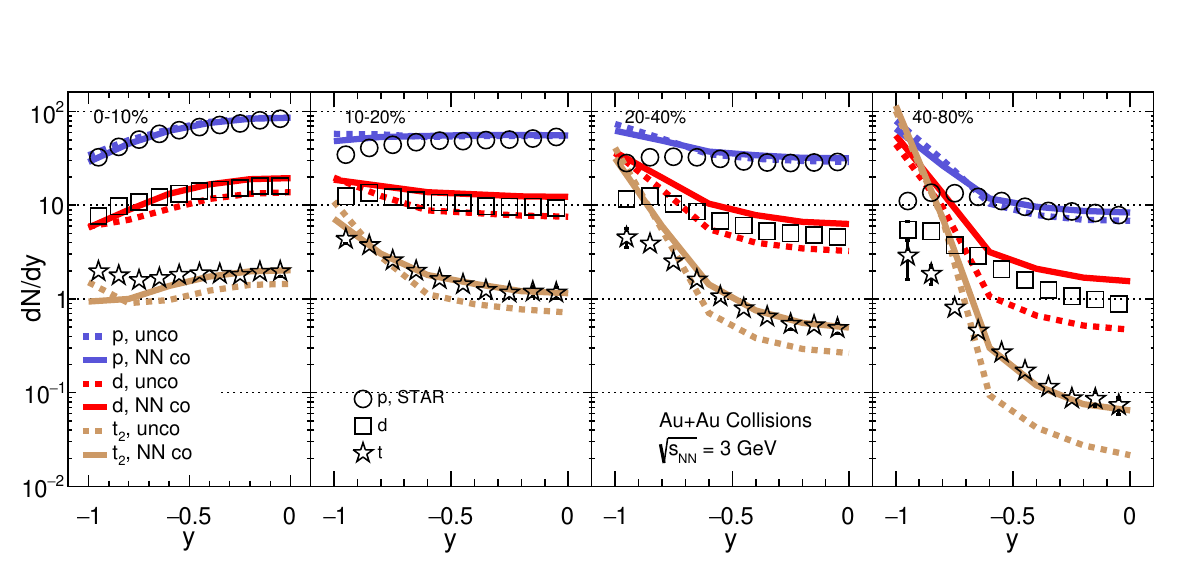}
\caption{Same as Fig.\ref{fig:dndy}, but with centrality bins defined by impact parameter.}
\label{fig:dndy_cen_by_b}
\end{figure*}

Figure \ref{fig:dndy_cen_by_b} illustrates the rapidity distribution of light nuclei yields in comparison with the STAR experiment in four different centralities for Au+Au collisions at $\sqrt{s_{NN}}$ = 3 GeV. The dashed lines(blue, red and brown) represent the distribution of proton, deuteron and triton yields, respectively, for two different initial configurations in the SMASH model: one with nucleon-nucleon correlation (solid lines) and one without (dashed lines).
In addition, the experimental data for proton, deuteron and triton yields are depicted using circles, squares and pentagons, respectively. 
As shown in Fig.~\ref{fig:dndy_cen_by_b}, 
the division of centrality based on impact parameter leads to a noticeable variation in particle yields. Specifically, the proton yield with correlation included actually increases with increasing centrality, which is contrary to the results obtained by dividing centrality using the multiplicity of final-state charged particles. The yield differences between deuteron and triton when centrality is divided by impact parameter are also larger than those when divided by the multiplicity of final-state charged particles. Therefore, the method of dividing centrality has a significant impact on particle yields that cannot be overlooked.

\subsection{Model Uncertainty}

\begin{figure*}[htp]
\centering
\includegraphics[width=1\textwidth,trim=10 5 8 20,clip]{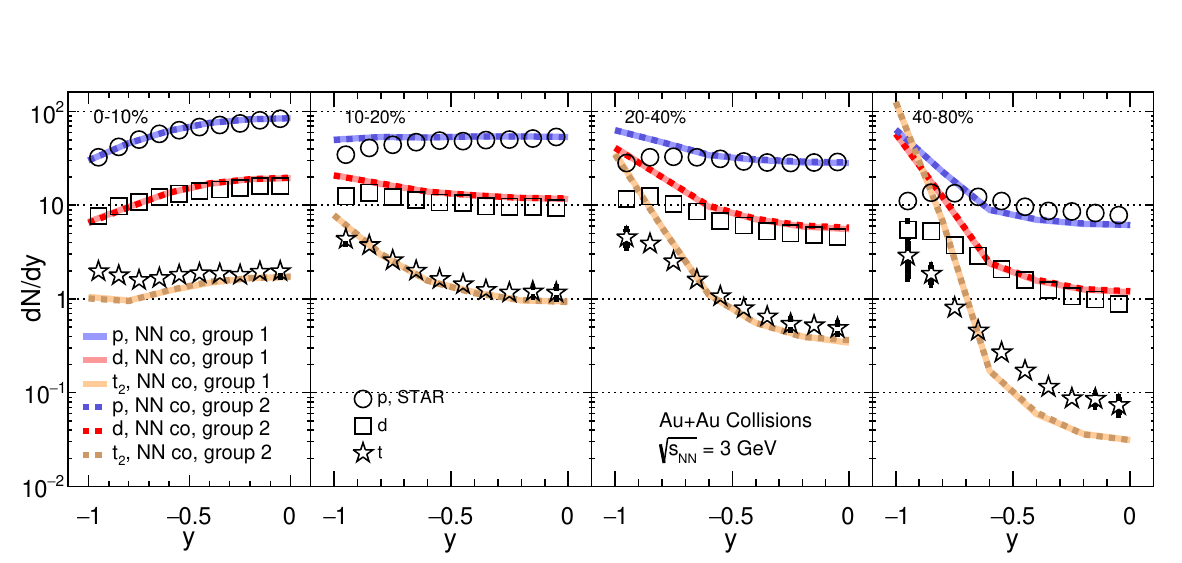}
\caption{Same as Fig.\ref{fig:dndy}, but calculated separately for two groups with nucleon-nucleon correlations, where each group contains 200,000 events.}
\label{fig:dndy_vs_2}
\end{figure*}

Figure \ref{fig:dndy_vs_2} illustrates the impact of statistical fluctuations derived from SMASH simulations. In this analysis, we compare two groups of light nuclei yields, each comprising 200,000 events, with nucleon-nucleon correlations taken into account. The solid and dashed lines represent group 1 and group 2, respectively. The results from these two groups exhibit substantial overlap, demonstrating that the current conclusions are not influenced by statistical fluctuations.

\subsection{Spectator Effect}

\begin{figure*}[htp]
\centering
\includegraphics[width=1\textwidth,trim=10 5 8 20,clip]{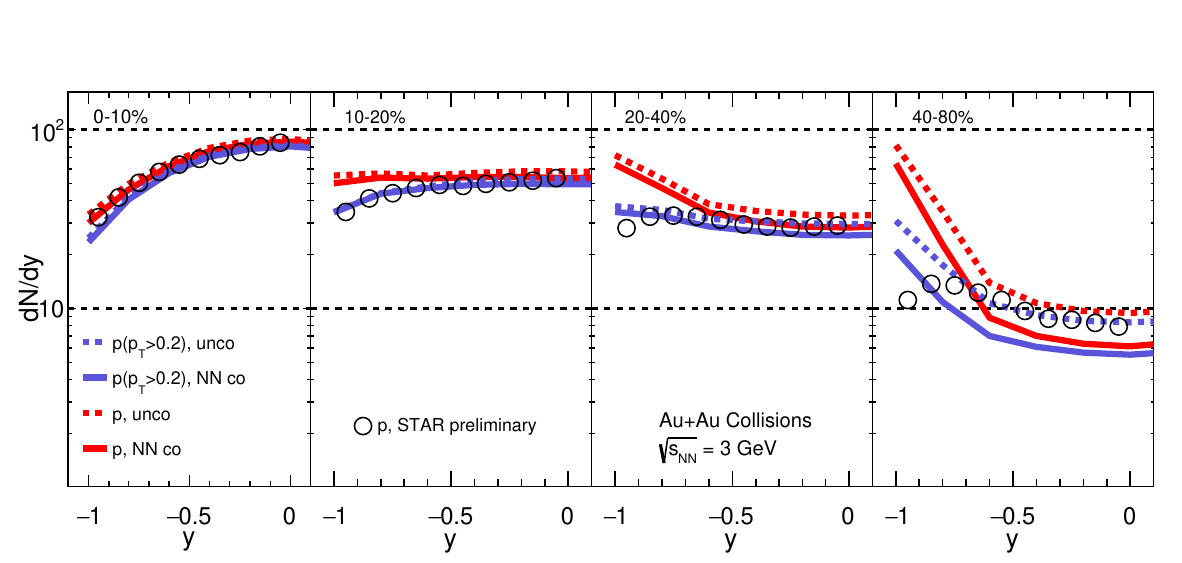}
\caption {(Color online) The yield of proton as a function of rapidity $y$ in 0$\sim$10\%, 10$\sim$20\%, 20$\sim$40\% and 40$\sim$80\% centrality using the SMASH model and coalescence method for Au+Au collisions at $\sqrt{s_{NN}}$ = 3 GeV. The dashed lines show the results for the initial nuclear structure without NN correlation in the initial state, and the solid lines show the results with NN correlation. Blue lines represent the yield of proton with $\eta$ and $p_T$ cut, while red lines represent the yield of proton without any cuts. The dots in each figure are the STAR experimental results\cite{10.21468/SciPostPhysProc.10.040,STAR:2023uxk}.
}
\label{fig:dndy_p_vs}
\end{figure*}

Figure \ref{fig:dndy_p_vs} shows the effect of spectator nucleons on the yield of proton as a function of rapidity $y$ across 0$\sim$10\%, 10$\sim$20\%, 20$\sim$40\% and 40$\sim$80\% centrality using the SMASH model and coalescence method for Au+Au collisions at $\sqrt{s_{NN}}$ = 3 GeV. The dashed lines show the results for the initial nuclear structure without NN correlation in the initial state, and the solid lines show the results with NN correlation. Blue lines represent the yield of proton with $p_T$ cut, while red lines represent the yield of proton without any cut. The dots in each figure are the STAR experimental results. 
Due to the small transverse momentum of spectators, we applied $p_T$ cut, using protons as an example. The resulting blue lines with cuts show a significant suppression at high rapidity compared to the red lines without cut. This confirms the impact of spectators on the light nuclei yield in the high rapidity region.

\subsection{Without Mean Field Potential}

\begin{figure}[htp]
\centering
\includegraphics[width=0.48\textwidth,trim=6 3 50 45,clip]{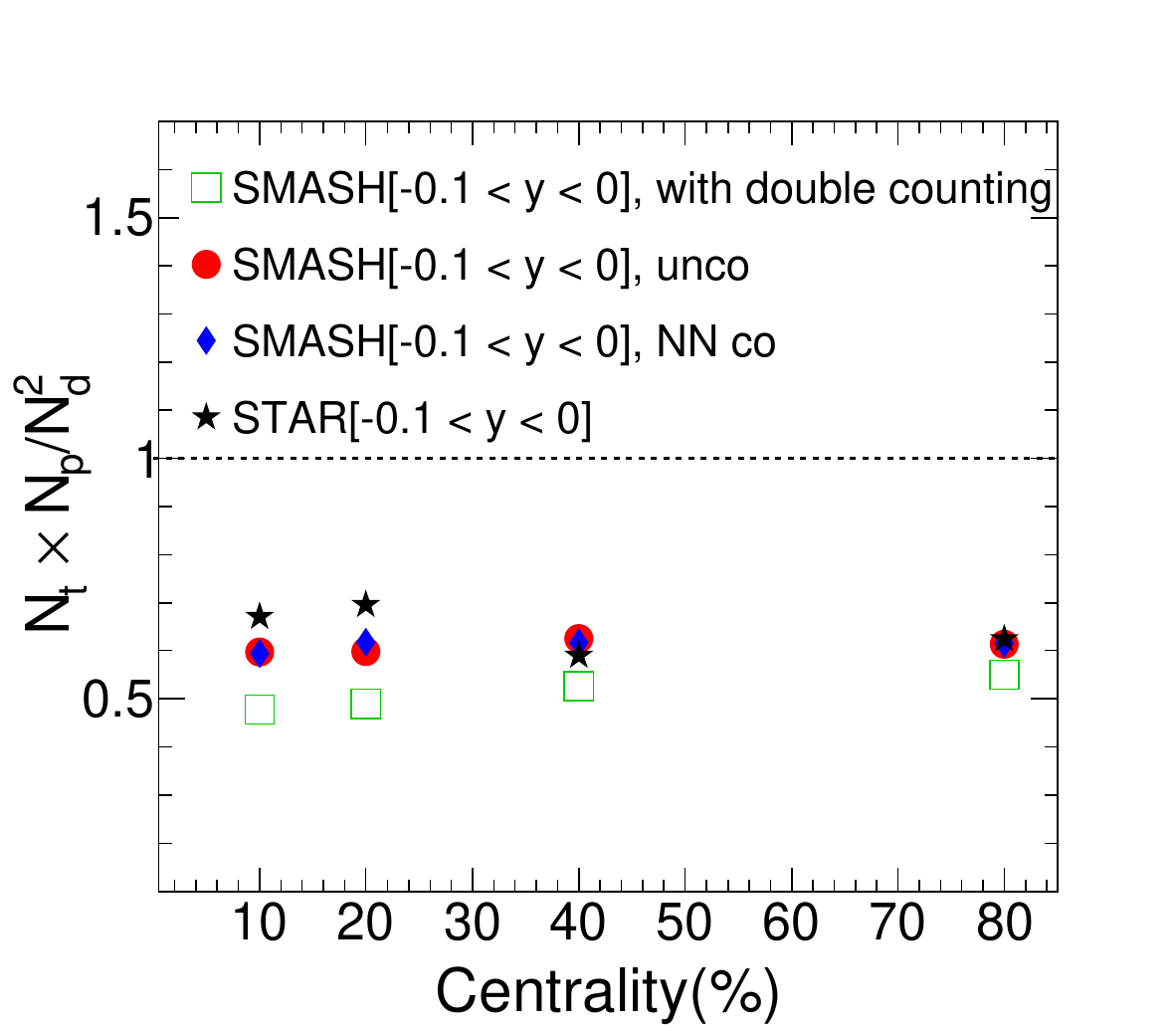}
\caption {As Fig.~\ref{fig:ratio_vs_cent}, but presented in cascade mode.}
\label{fig:ratio_vs_cent_cas}
\end{figure}

\begin{figure*}[htp]
\centering
\includegraphics[width=1\textwidth,trim=10 5 8 20,clip]{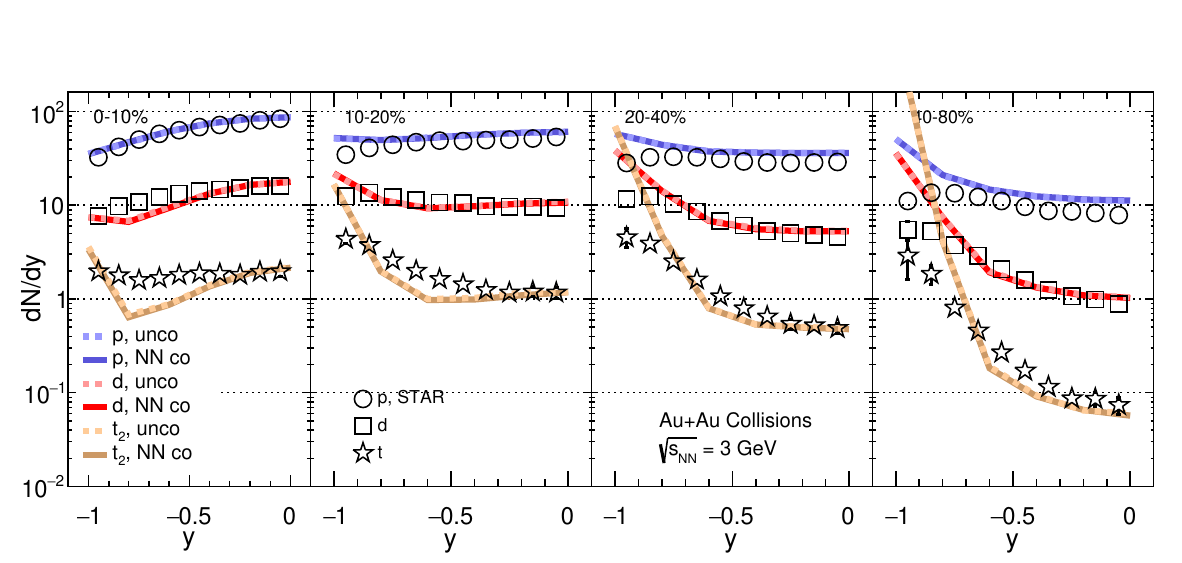}
\caption {As Fig.~\ref{fig:dndy}, but presented in cascade mode.}
\label{fig:dndy_cas}
\end{figure*}

Figure ~\ref{fig:ratio_vs_cent_cas}, analogous to Fig.~\ref{fig:ratio_vs_cent}, depicts the double ratio in cascade mode, where effect of NN correlation is negligible.  The same scenario is observed in Figure \ref{fig:dndy_cas}, analogous to Fig.~\ref{fig:dndy}, which demonstrate that, in cascade mode, results with NN correlations brings a negligible impact on particle yields. In contrast, the mean field mode, as illustrated in Fig.~\ref{fig:dndy}, exhibits a significant discrepancy. This is attributable to the mean field mode's inclusion of multiple particle interactions, which can influence the yield of light nuclei. Given the low collision energy, the mean field model is preferred for its realistic representation of the collision conditions.

\bibliography{ref}

\end{document}